\documentclass[12pt]{article}

\pdfoutput=1
\usepackage{jheppub}
\usepackage{graphicx,epsfig,amsmath,amssymb,booktabs}
\usepackage{caption}
\usepackage{subcaption}

\title{Diphoton production in association with two bottom jets}

\author{Daniel F\"ah,}
\author{Nicolas Greiner}
 
\affiliation{Physik Institut, Universit{\"a}t Z{\"u}rich, Winterthurerstr.190, 8057 Z\"urich,Switzerland}

\preprint{{\small  ZU-TH 19/17}}

\abstract{
 We study the production of a photon pair in association with two bottom jets at the LHC. This process constitutes
 an important background to double Higgs production with the subsequent decay of the two Higgs bosons into a pair of photons
 and b-quarks respectively. We calculate this process at next-to-leading order accuracy in QCD and find that 
 QCD corrections lead to a 
 substantial  increase of the production cross section due to new channels opening up at next-to-leading order and
 their inclusion is therefore
 inevitable for a reliable prediction. 
 Furthermore, the approximation of massless b-quarks is scrutinized by calculating the
 process with both massless and massive b-quarks. We find that the massive bottom quark leads to a substantial reduction of the
 cross section where the biggest effect is however due to the use of a four flavor PDF set and the corresponding smaller values
 for the strong coupling constant.
}


\begin{document}

\maketitle

\section{Introduction}
The boson discovered at the LHC \cite{Aad:2012tfa,Chatrchyan:2012ufa} seems to be in very good agreement
with prescription of a Standard Model like Higgs boson. In the Standard Model the Higgs mass is the 
only free parameter in the theory and its precise determination was one of the main experimental
targets \cite{Aad:2015zhl}. Furthermore, the Standard Model predicts the shape of the Higgs potential,
so a measurement of the parameters of the potential will allow us to discriminate a Standard Model Higgs
boson from various BSM scenarios. This however requires the measurement of the Higgs self coupling, which
can be measured in Higgs boson pair production processes. The value of the Higgs mass allows for measurements in
a variety of decay channels and both ATLAS and CMS have performed studies to measure the Higgs self coupling, e.g. in
the decay channels $\gamma\gamma
b\bar{b}$~\cite{ATLAS-CONF-2016-004,Khachatryan:2016sey,Aad:2015xja,Aad:2014yja}, 
$b\bar{b}b\bar{b}$~\cite{Aaboud:2016xco,CMS:2016tlj,Aad:2015xja,Khachatryan:2015yea,Aad:2015uka}, 
$\gamma\gamma W W^*$, $b\bar{b}W W^*$,
$\tau^+\tau^-b\bar{b}$~\cite{ATLAS:2016qmt,CMS:2016cdj,CMS:2016ymn,CMS:2016rec,CMS-PAS-HIG-16-013,CMS:2016ugf,CMS:2016zxv,Aad:2015xja}.\\
From a Standard Model calculational point of view, the signal process (i.e. the production of a Higgs boson pair)
is known at leading order in the full theory \cite{Eboli:1987dy,Glover:1987nx,Plehn:1996wb}, and in  various approximations
taking higher order corrections into account \cite{Dawson:1998py,Maltoni:2014eza,Grigo:2013rya,Grigo:2014jma,Grigo:2015dia,Degrassi:2016vss,
deFlorian:2013uza,deFlorian:2013jea,Shao:2013bz,deFlorian:2015moa}. Only very recently the NLO result taking full top
mass dependence into account became available \cite{Borowka:2016ehy,Borowka:2016ypz}.\\
In this paper we focus on one possible decay channels, namely where one Higgs decays into a pair of photons, whereas the 
second decays into a pair of b-quarks. This process can be seen as a compromise between a four $b$-quark signal and 
a four photon signature. The first would benefit from a large $H\to b \bar{b}$ branching ratio but suffers from a large 
irreducible QCD background, whereas the latter exhibits a very clean signal with four photons, but suffers from a very small 
$H\to \gamma \gamma$ branching ratio.\\
In the case of massless $b$-quarks the process $\gamma \gamma b \bar{b}$ can be seen as a subset of the process $\gamma \gamma jj$ which 
is known at NLO in QCD \cite{Gehrmann:2013bga,Badger:2013ava,Bern:2014vza}. The main motivation for the general two jet process 
was however more to assess the background of a single Higgs in VBF production rather than focusing on final state b-quarks.
As we will see, the tagging of two final state b-jets significantly alters the behavior of the higher order corrections and therefore
this process cannot be directly compared to the general two jet process.\\
The paper is organized as follows. In section \ref{sec:setup} we describe the setup that has been used to obtain the numerical results
which we discuss in section \ref{sec:results}. Finally we conclude in section \ref{sec:conclusions}.

\section{Calculational setup}
\label{sec:setup}
The  NLO corrections are calculated by combining the two automated programs \textsc{GoSam}
\cite{Cullen:2011ac,Cullen:2014yla} for the generation and evaluation
of the virtual one-loop amplitudes, and the Monte Carlo event
generator \textsc{Sherpa} \cite{Gleisberg:2008ta}. The combination between the two is realized
using the standardized Binoth Les Houches Accord
\cite{Binoth:2010xt,Alioli:2013nda}.\\
\textsc{GoSam} is based on an algebraic approach where
$d$-dimensional integrands are generated using Feynman diagrams. It
uses \textsc{QGraf}~\cite{Nogueira:1991ex} and
\textsc{Form}~\cite{Vermaseren:2000nd,Kuipers:2012rf} for the diagram
generation, and \textsc{Spinney}~\cite{Cullen:2010jv}
and \textsc{Form} to write an
optimized Fortran output. For the reduction of the tensor integrals
we used \textsc{Ninja}~\cite{Mastrolia:2012bu,vanDeurzen:2013saa,Peraro:2014cba},
which carries out the reduction on the integrand level in a fully automated way via
Laurent expansion. Alternatively one can choose other reduction strategies such as OPP reduction
method~\cite{Ossola:2006us,Mastrolia:2008jb,Ossola:2008xq} which is
implemented in \textsc{Samurai}~\cite{Mastrolia:2010nb} or methods based on
tensor integral reduction as implemented in
\textsc{Golem95}~\cite{Heinrich:2010ax,Binoth:2008uq,Cullen:2011kv,Guillet:2013msa}.
For the evaluation of the remaining scalar integrals we have used
\textsc{OneLoop}~\cite{vanHameren:2010cp}.\\
The evaluation of all tree-level like matrix elements within \textsc{Sherpa} has been performed
using \textsc{Comix} \cite{Gleisberg:2008fv}, the subtraction terms have been
calculated with the \textsc{Sherpa}'s implementation of the Catani-Seymour dipole formalism 
\cite{Catani:1996vz,Catani:2002hc}.

\section{Numerical results}
\label{sec:results}
In the following we present numerical results for the LHC with a center of mass energy of $\sqrt{s}=13 \text{TeV}$. 
To assess $b$-mass effects the calculation has been carried out with both massless $b$-quarks in the 5 flavor
scheme as well as with massive $b$-quarks in the 4 flavor scheme.
\subsection{Cuts and parameter settings}
For the massless case we have used the CT10nlo pdf set \cite{Lai:2010vv}  and the CT10nlo\_nf4 set for the massive case respectively. In the massive case the the $b$-mass has been set to $4.7$ GeV. \\
Renormalization- and factorization scales are set to be equal and the central scale was chosen to be
\begin{equation}
 \mu_R = \mu_F =\frac{1}{2}\sqrt{m^2_{\gamma \gamma} +\left( \sum_i p_{T,i}\right)^2}\;,
\end{equation}
where the sum runs over the final state partons.
As this process contains external photons the electroweak coupling constant $\alpha$ is set to $\alpha = 1.0/137.03599976$.
We have included top-quark loops in the virtual corrections with a top mass of $m_t = 171.2 $ GeV.\\
The presence of final state photons requires the application of a photon isolation criterion to render the 
NLO corrections finite. We employed a smooth cone isolation criterion \cite{Frixione:1998jh} with the following parameters:
\begin{equation}
 R=0.4,\quad \epsilon=0.05,\quad n=1\;.
\end{equation}
Additionally the isolated photons are required to fulfill
\begin{equation}
 p_{T,\gamma} > 30\; \text{GeV},\quad |\eta_{\gamma}|<2.5\;.
\end{equation}
The QCD partons are clustered with an anti-$k_T$ algorithm \cite{Cacciari:2008gp} contained in the Fastjet package \cite{Cacciari:2011ma}. 
The jet radius has been set to $R=0.4$ and events where both $b$-quarks are clustered into a jet are rejected in order 
to ensure that there are at least two $b$-jets present in the final state. For the jets we require 
\begin{equation}
 p_{T,j} > 20\; \text{GeV}, \quad |y_j|<4.4\;.
\end{equation}

\label{sec:settings}
\subsection{Cross sections and differential distributions}
\label{sec:xs}
\begin{figure}[t!]
  \centering
  \includegraphics[width=0.8\textwidth]{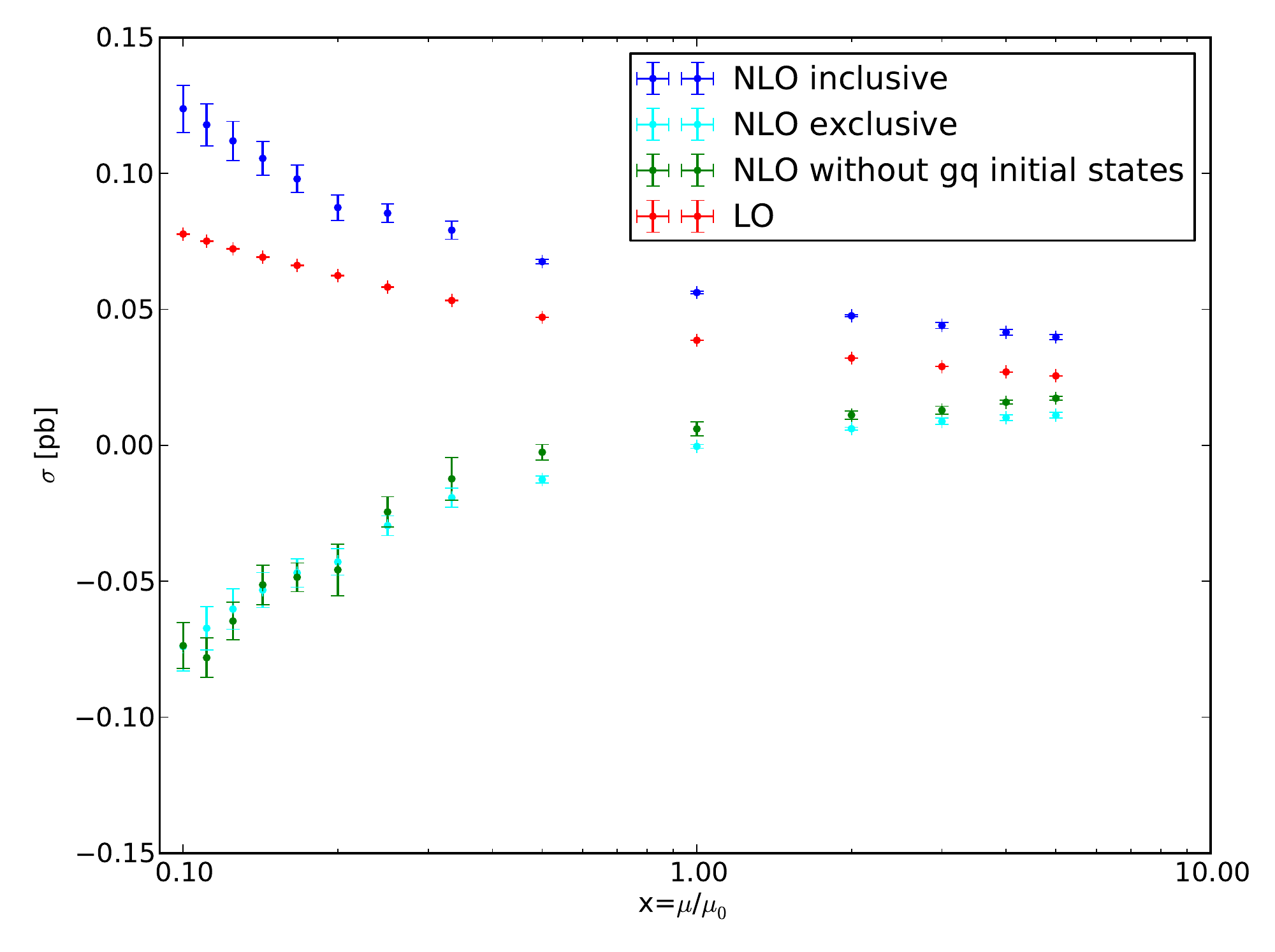}
  \caption{\label{fig:xsec} Total cross sections at LO and NLO for massless $b$-quarks. In addition it shows the NLO
  cross section where the initial $qg$-channel has been removed as well as the exclusive two $b$-jet cross section 
  where a veto on a third jet has been imposed.}
\end{figure}

\begin{figure}[t!]
  \centering
  \includegraphics[width=0.49\textwidth]{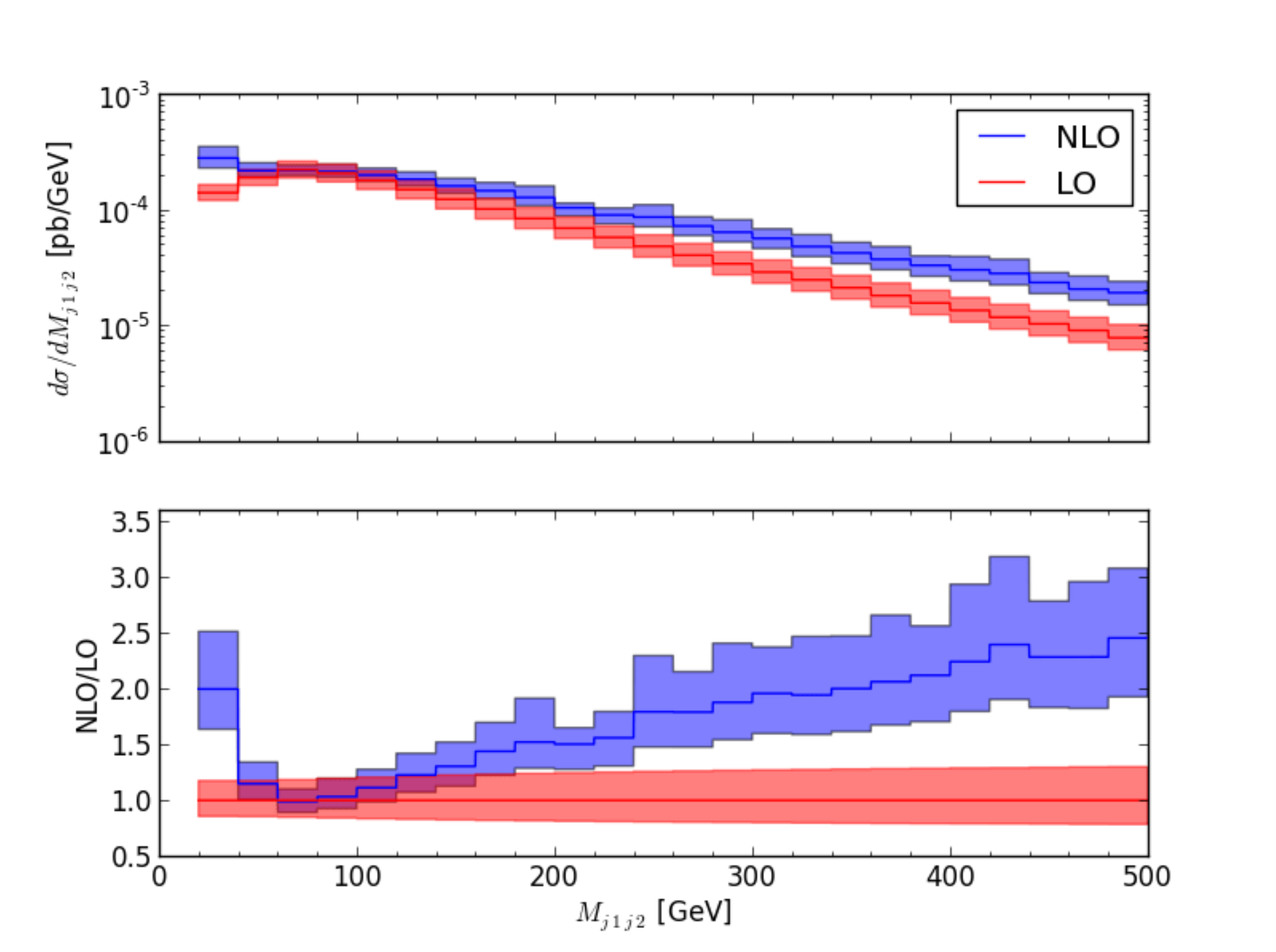}
  \hfill
  \includegraphics[width=0.49\textwidth]{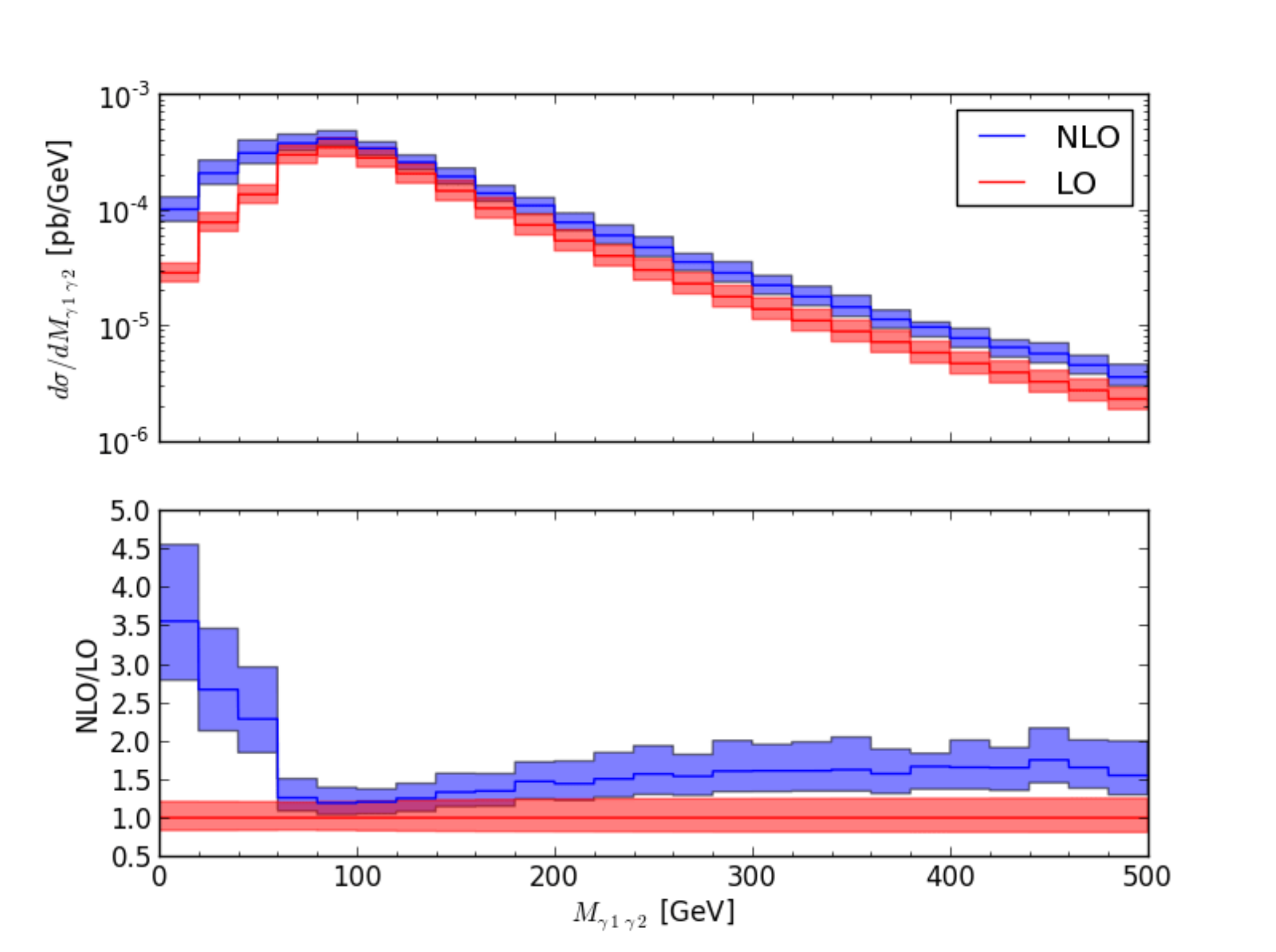}\\
  \includegraphics[width=0.49\textwidth]{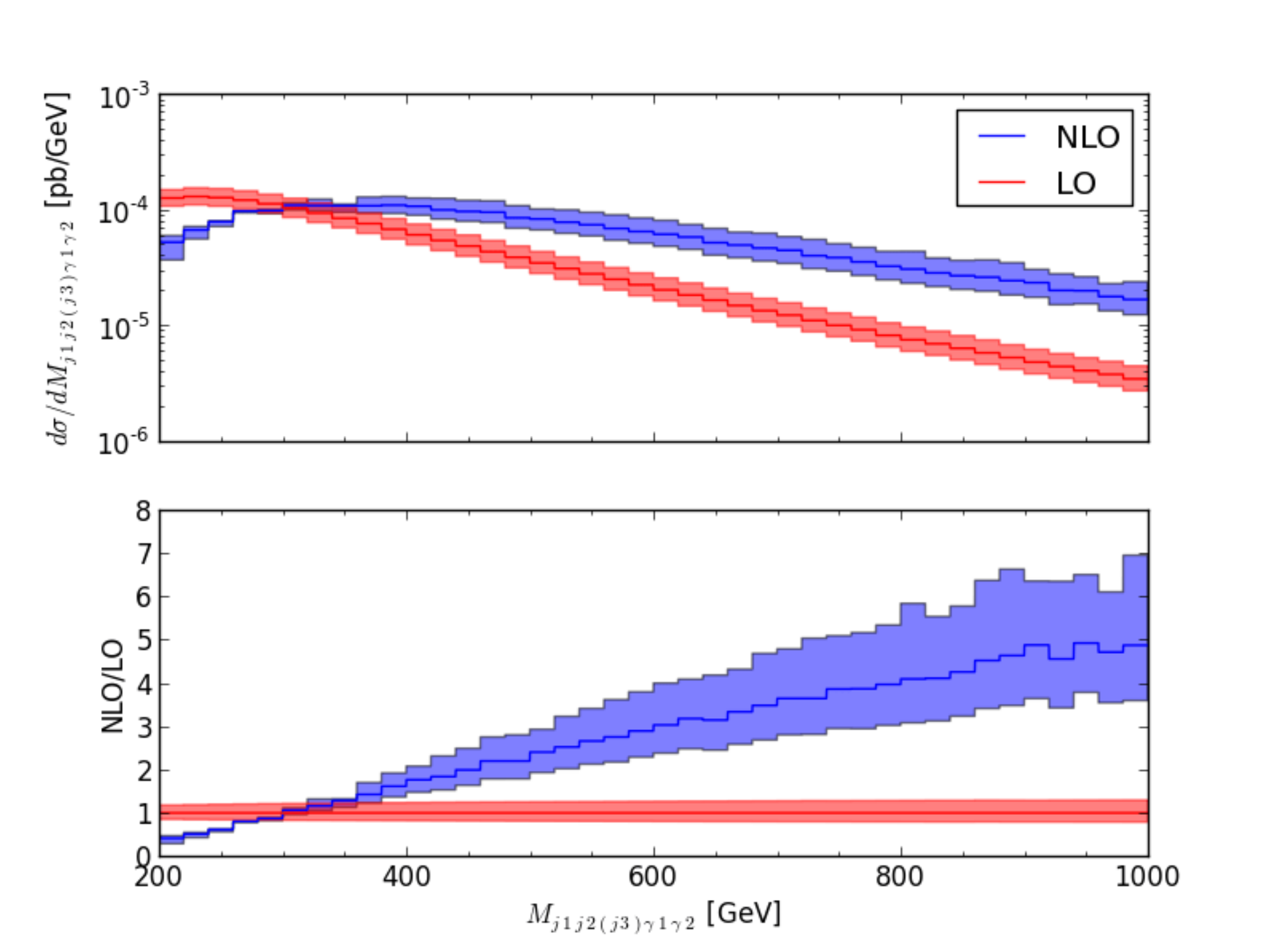}
  \caption{\label{fig:m_invar} Invariant mass distributions of the two leading jets (upper left), the two photons (upper right)
  and the total invariant mass.}
\end{figure}

\begin{figure}[t!]
  \centering
  \includegraphics[width=0.49\textwidth]{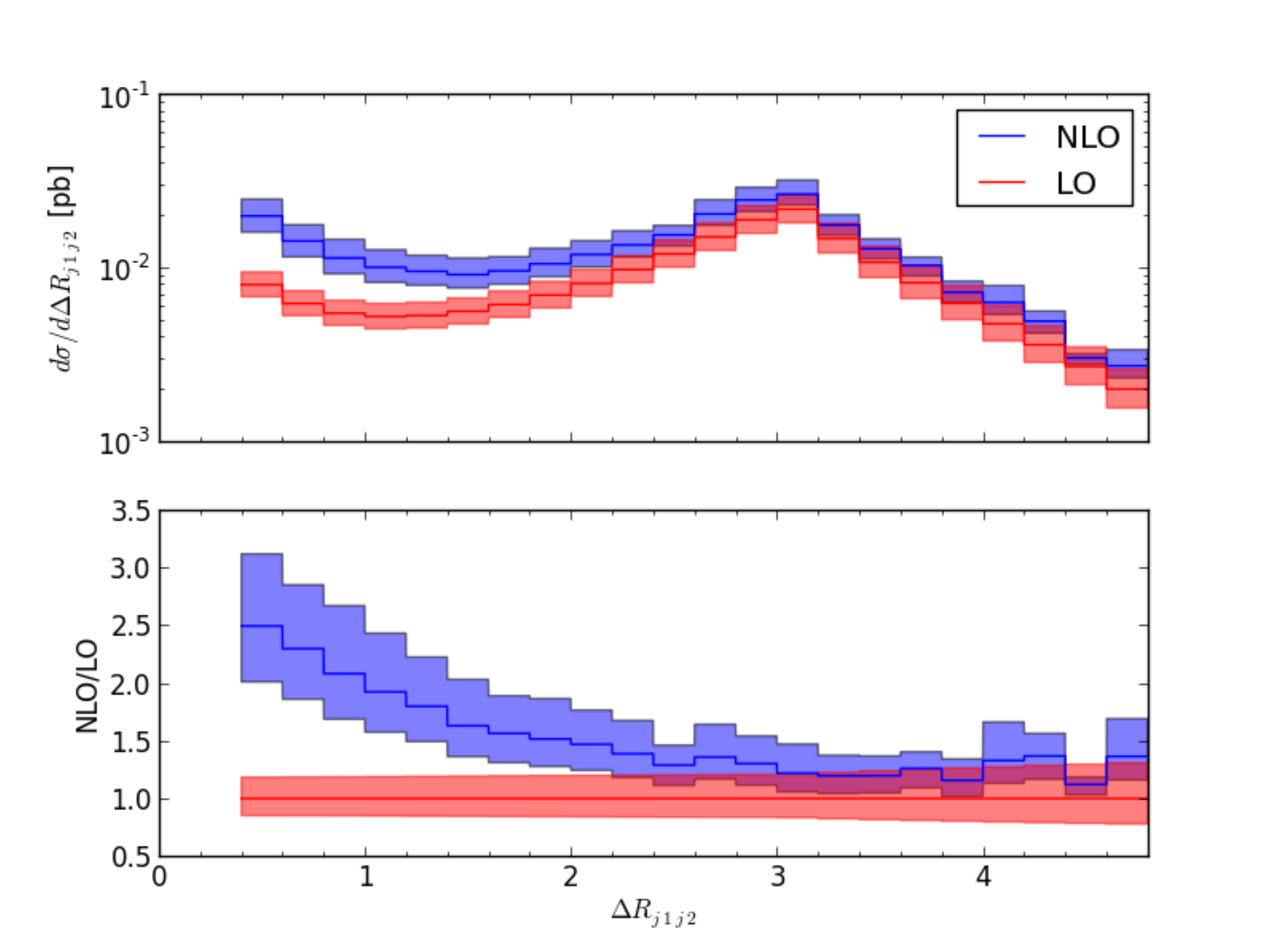}
  \hfill
  \includegraphics[width=0.49\textwidth]{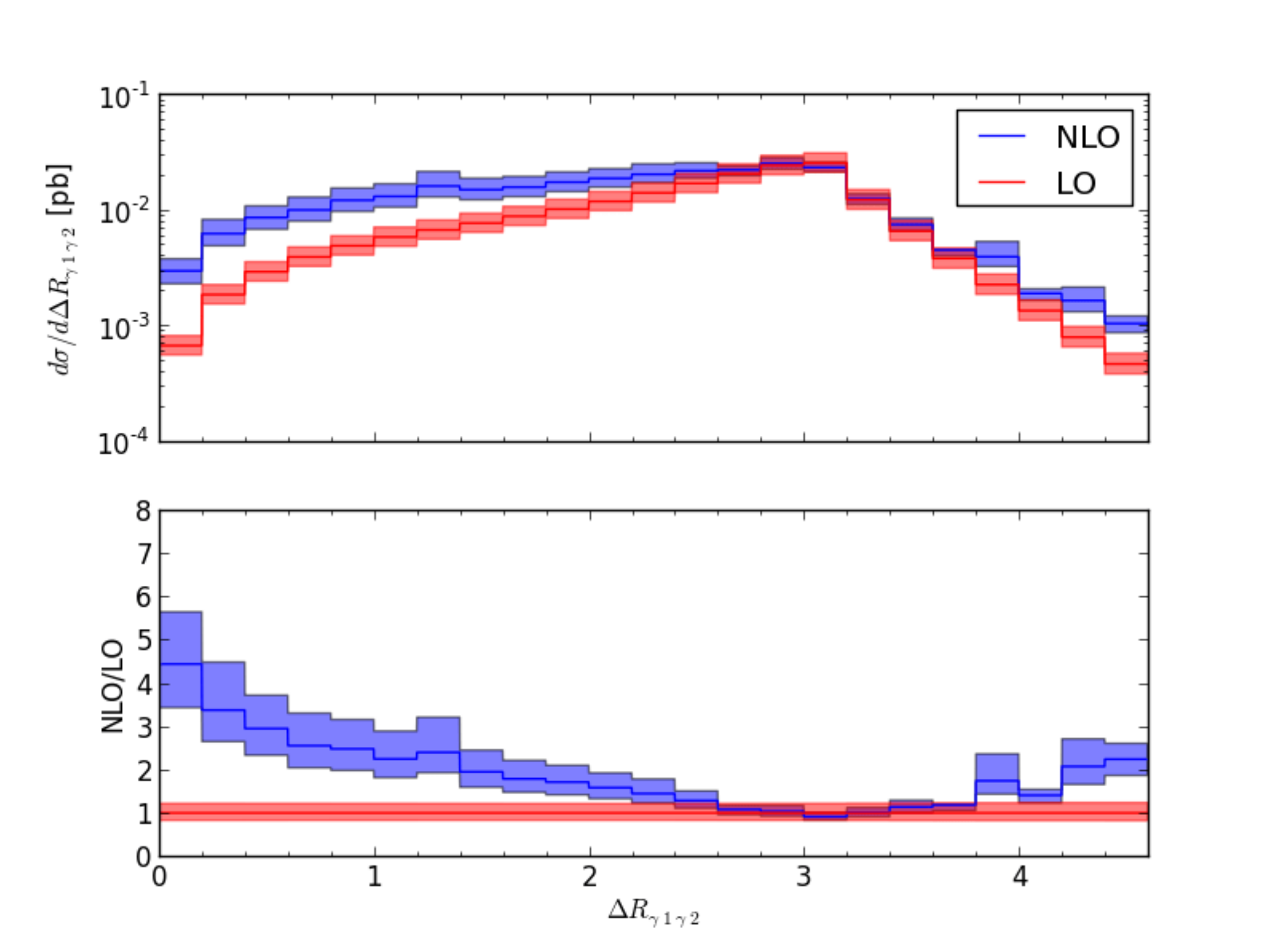}
  \caption{\label{fig:dr} $R$-separation of the two leading jets (l.h.s) and the two photons (r.h.s)}
\end{figure}

\begin{figure}[t!]
  \centering
  \includegraphics[width=0.49\textwidth]{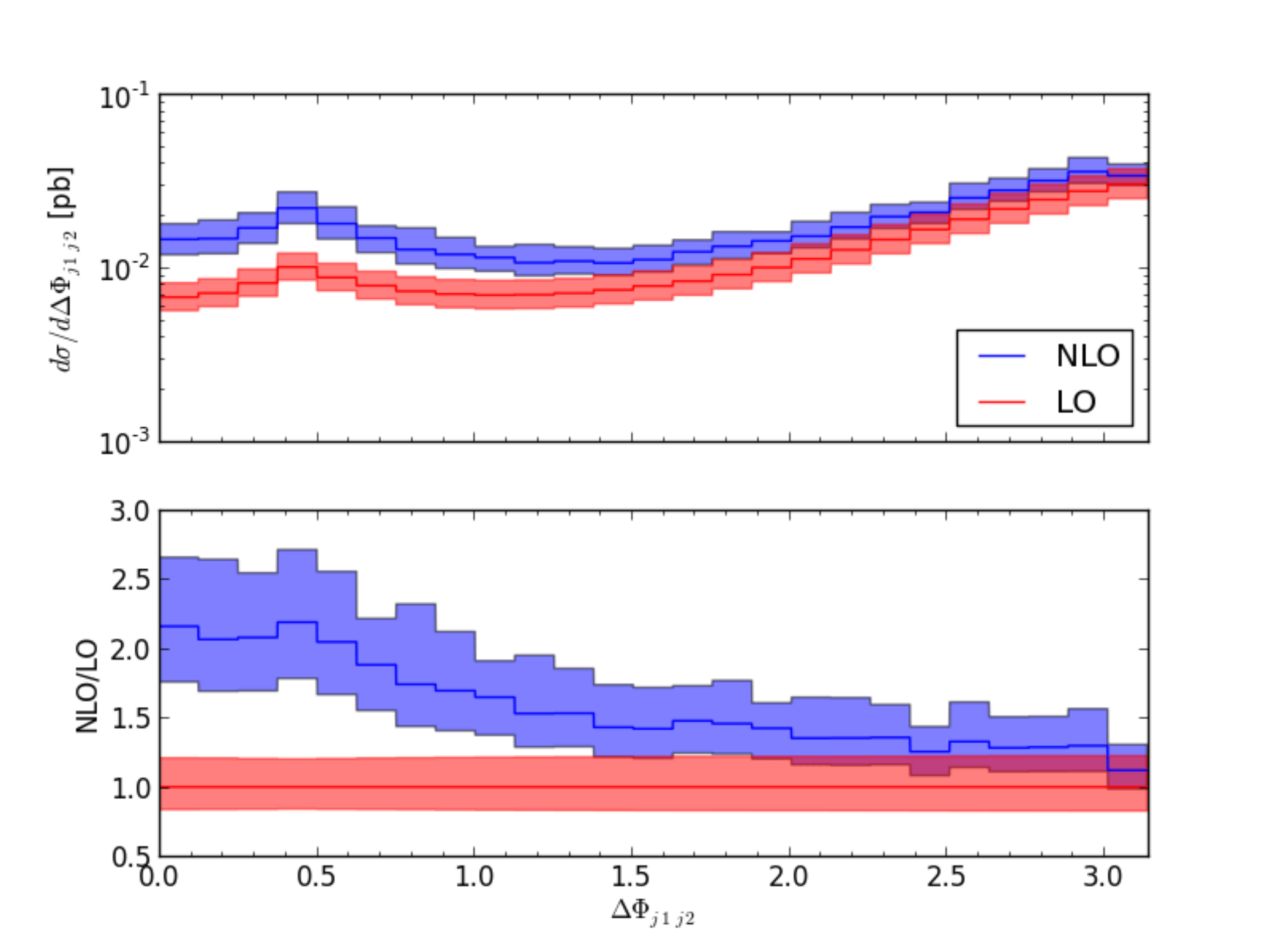}
  \hfill
  \includegraphics[width=0.49\textwidth]{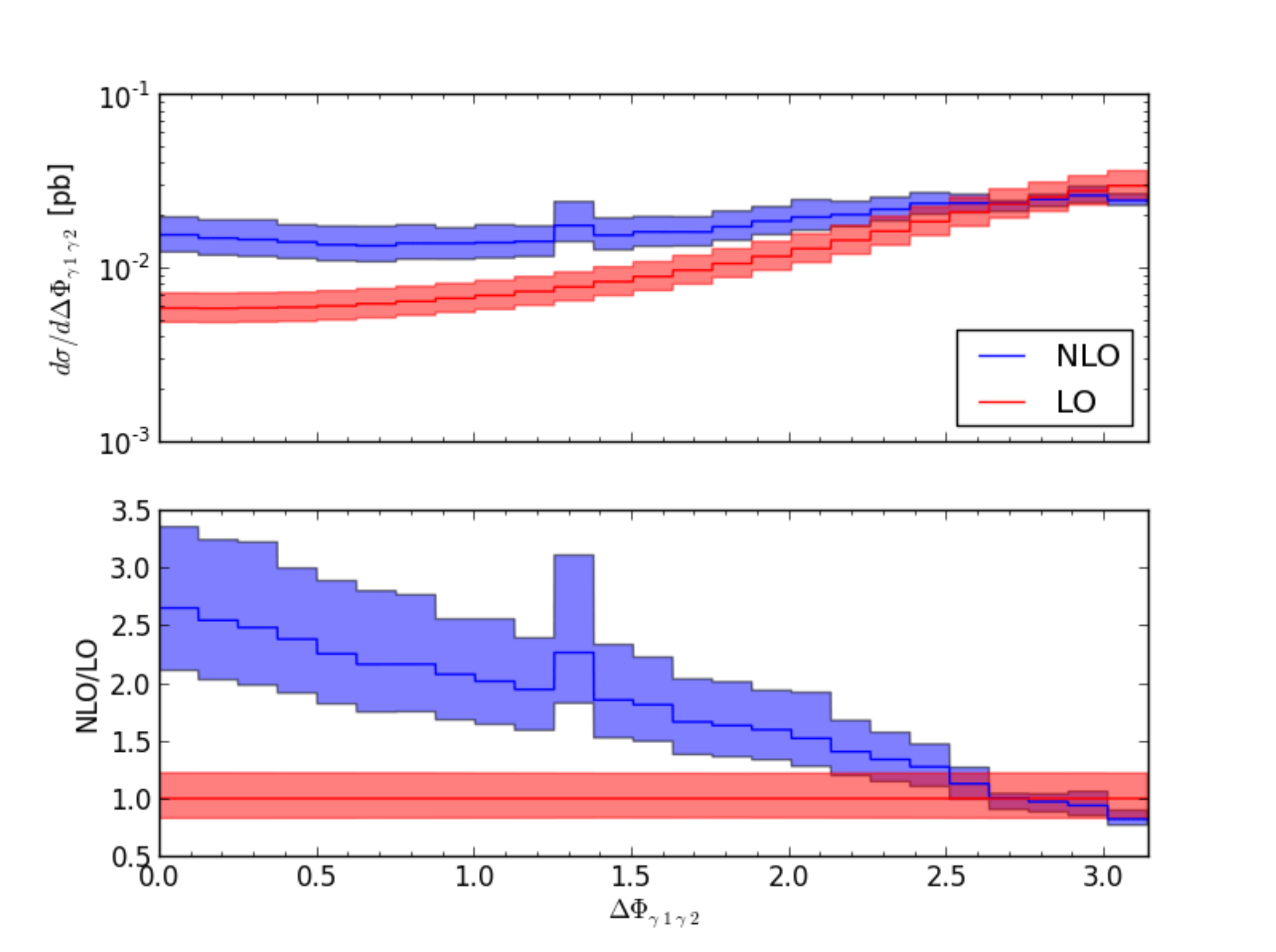}
  \caption{\label{fig:dphi} Azimuthal separation of the two leading jets (l.h.s) and the two photons (r.h.s)}
\end{figure}

We start the discussion of the numerical results with the case of a massless $b$-quark. We assess the theoretical uncertainty by 
usual scale variation of a factor of two around the central scale. Based on the cuts and settings described above we find for the 
total cross section 
\begin{equation}
\label{xsec:massless}
 \sigma_{LO} = 38.6^{+22\%}_{-17 \%}\; \text{fb},\quad \sigma_{NLO} = 56.2^{+20\%}_{-15\%}\;.
\end{equation}
From Eq. \ref{xsec:massless} one can see that the NLO corrections enhance the total cross section by almost fifty per cent.
It also shows that the theoretical uncertainty does not improve at NLO, instead for both LO and NLO one obtains an uncertainty of 
$15-20$\% in each direction when varying the scale by a factor of two. This situation is shown more explicitly in Fig.~\ref{fig:xsec}
where we show the cross section as a function of the scale for a broader range.
Looking at the curve for the inclusive NLO result one sees a born-like behavior even at NLO. In particular no reduction 
on the scale dependence is obtained throughout the whole range of scales. The typical turnover that one expects at NLO is 
not present.\\
A special feature of this process is that the leading order process is mediated by two types of initial state, the $q\bar{q}$- and
the $gg$-channel. At NLO however also the quark-gluon channel is opening up in the real emission. In order to investigate whether 
it is this channel that is responsible for the tree-level like behavior we made two different checks. First, we completely
remove the quark-gluon channel from the process (green curve). This has a tremendous impact on the NLO result rendering the 
corrections negative over the whole range and the absolute value increases when going to smaller scales which even leads to unphysical
negative cross sections for scales smaller than the central scale. Removing a production 
channel is of course not a physical meaningful procedure but it shows that this channel is indeed responsible for the 
behavior of the inclusive NLO cross section. A physically well defined strategy however is to impose a jet veto on a possible 
third jet. A jet veto effectively cuts away an intrinsically positive contribution from the real emission and will therefore
lead to a decrease of the NLO result. The exclusive NLO result is given by the turquois curve. Interestingly the two approaches
lead to very similar results. Even though vetoing a jet is a well defined procedure it leads to negative cross sections for 
scales smaller than the central scale. This indicates that the central scale could be chosen to be larger although it has been
proven to be a good choice for the general diphoton plus two jets process \cite{Gehrmann:2013bga}. It is clear that imposing
a jet veto raises the question to what extent possible resummation effects can change the result and the associated
theoretical uncertainty. This is however beyond the scope of this paper. The results show that this process is highly 
sensitive to an additional jet veto and that the scale variation might therefore not be an accurate measure of the 
theoretical uncertainty. The inclusive NLO result seems however suitable as a conservative estimation.\\
We now turn to the discussion of the differential distributions. 
\begin{figure}[t!]
  \centering
  \includegraphics[width=0.49\textwidth]{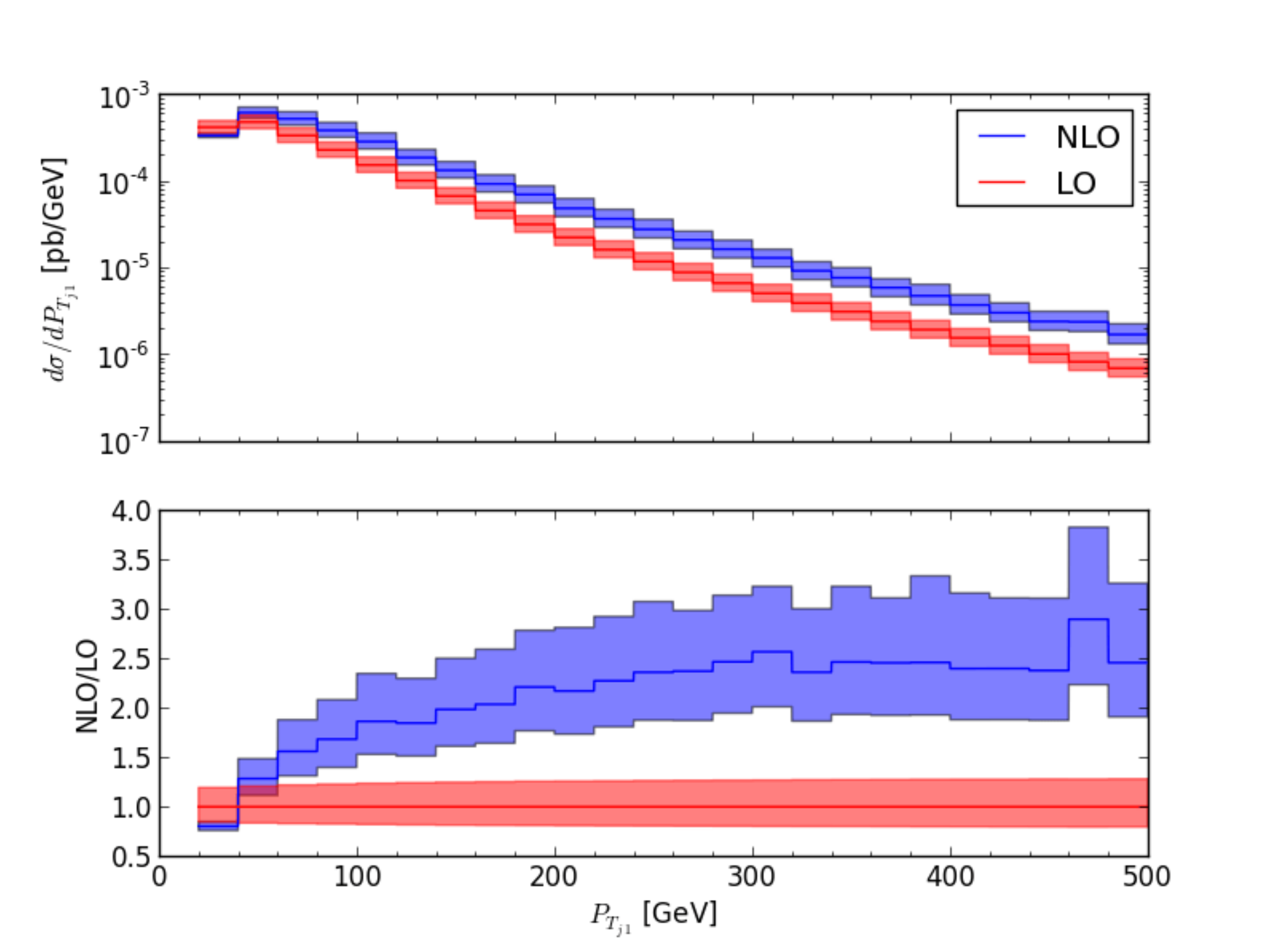}
  \hfill
  \includegraphics[width=0.49\textwidth]{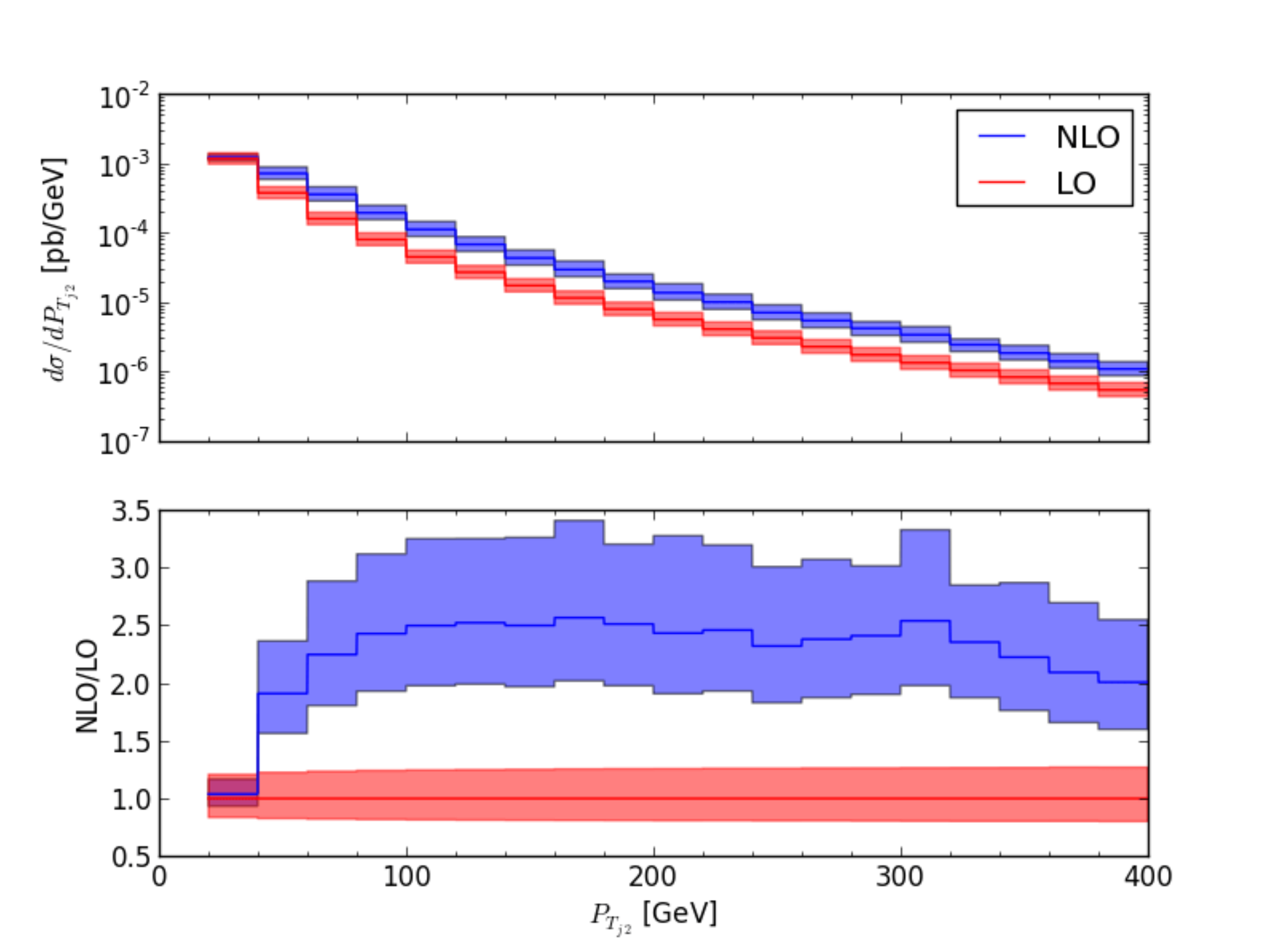}
  \caption{\label{fig:pt_jet} Transverse momentum of the two leading $b$ - jets.}
\end{figure}
Fig.~\ref{fig:pt_jet} shows the $p_T$ distribution of the two leading $b$-jets, where the jets are $p_T$-ordered.
For both jets the NLO corrections for low values of $p_T$ are relatively small which means that the NLO result agrees with the 
leading order result within the systemic uncertainty.  Also the size of the NLO uncertainty is reduced compared to the
leading order uncertainty. Going higher in $p_T$ however very rapidly increases the NLO corrections and from the order of 
$100$ GeV on the differential $k$-factors are in the range of $2-2.5$. Also the size of the NLO uncertainty band increases and for values beyond $\sim 100$ GeV the uncertainties at NLO are roughly twice as big as the LO ones.\\
\begin{figure}[t!]
  \centering
  \includegraphics[width=0.49\textwidth]{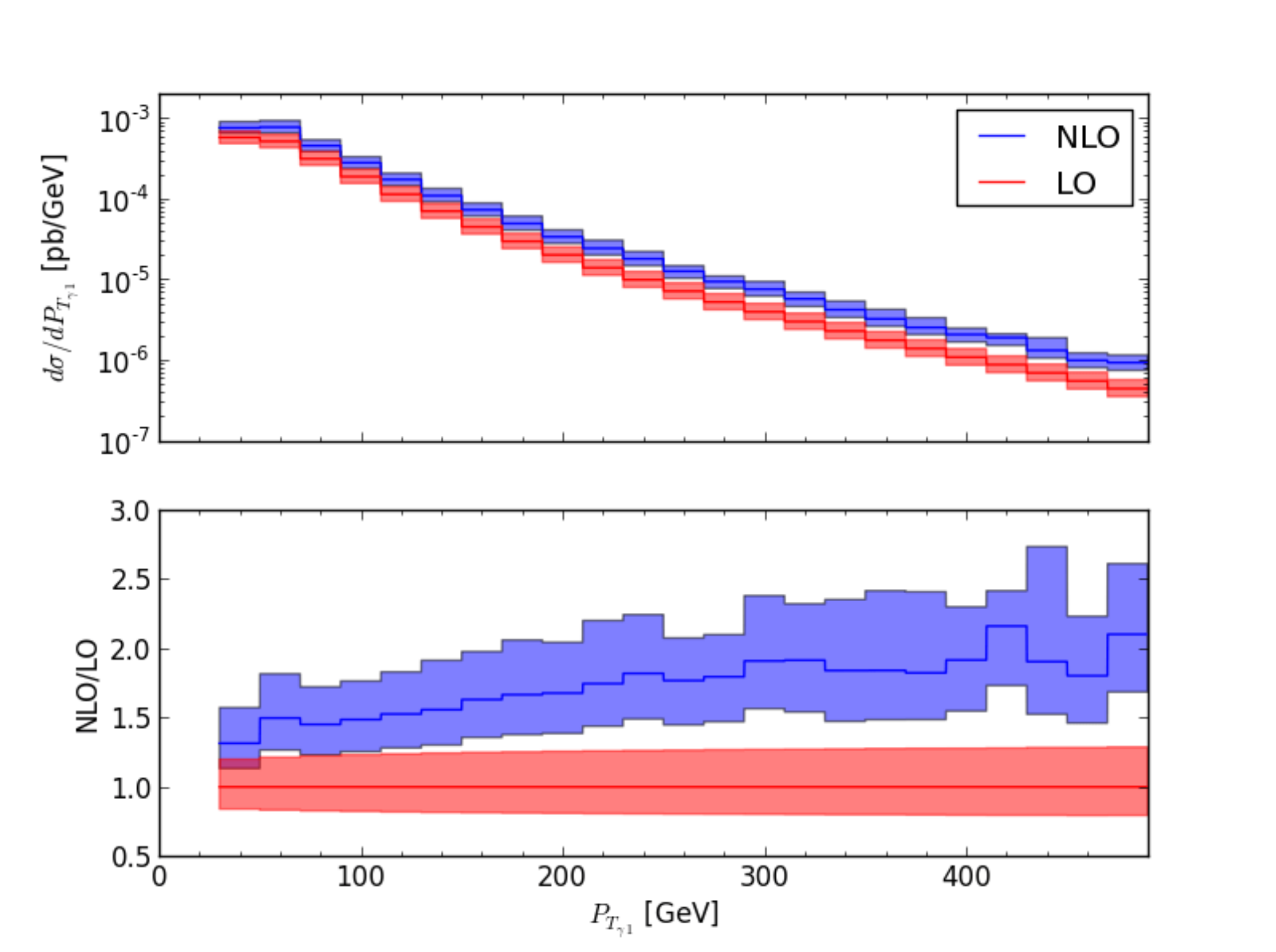}
  \hfill
  \includegraphics[width=0.49\textwidth]{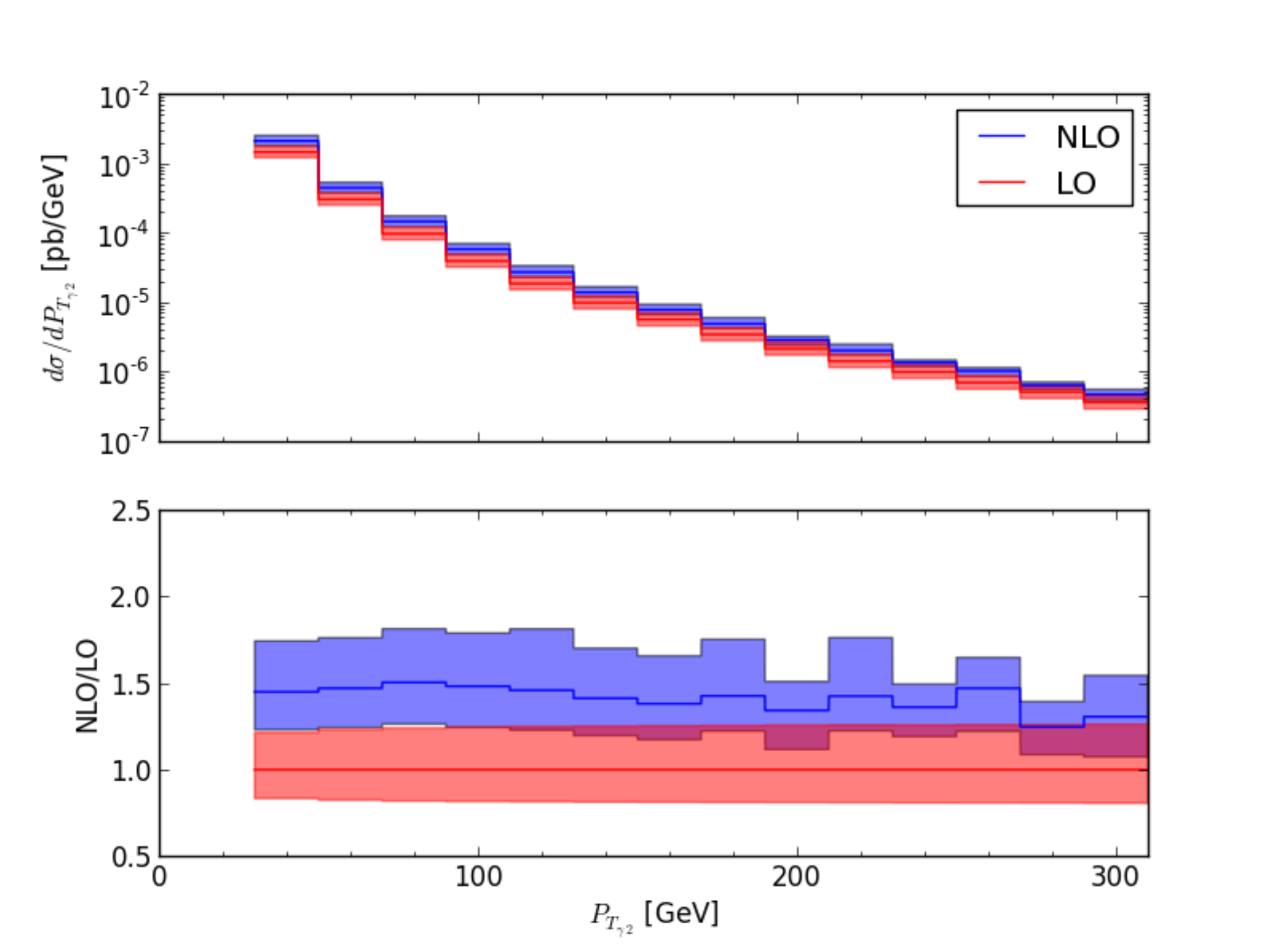}
  \caption{\label{fig:pt_photon} Transverse momentum of the two photons.}
\end{figure}
For the transverse momentum distribution of the photons shown in Fig.~\ref{fig:pt_photon} the behavior is less pronounced than
for the jets. For the leading photon the corrections are smallest for low values of $p_T$ and rise almost linearly with increasing
 transverse momentum leading to a $k$-factor of almost two for values around $500$ GeV. Similar to the jet distributions
 there is basically no overlap between the uncertainty bands. The subleading photon shows a milder behavior compared to
 the leading photon. Although the uncertainty bands also hardly overlap the differential $k$-factor is flat to a good approximation.
  For both photons one sees that the size of the NLO uncertainty is roughly of the same size as the LO uncertainties whereas
   for the jets the NLO uncertainties were larger except for small values of $p_T$.\\
As this process constitutes a background to double Higgs production, the invariant mass distributions are also essential. 
In the upper row of Fig.~\ref{fig:m_invar} we show the invariant masses of the two leading $b$- jets and of the two photons.
 In both cases one 
observes a significant shape distortion by the NLO corrections. They exhibit large corrections at low values followed by a
minimum in the range of $60-80$ GeV. In the case of the jets the NLO corrections then increase roughly linearly again, leading to
substantial corrections for invariant masses beyond say $200$ GeV. For the photons this behavior is mitigated and the 
differential $k$-factor is flat to a good approximation in the mass range beyond $200$ GeV. It is worth noting that in the range
around the Higgs mass the corrections are rather mild and one still finds an overlap between the uncertainty bands. And in 
particular the NLO behavior for low invariant masses allows to reduce the NLO corrections by imposing an appropriate cut 
around the Higgs mass. The plot in the lower row of Fig.~\ref{fig:m_invar} shows the total invariant mass of the final state where
the sum runs over the two photons and the jets. There we see a drastic change in the shape of the distribution when going
from leading order to next-to-leading order.  At low invariant masses the NLO corrections are negativ and substantial but
then increase linearly and lead to substantial positive corrections in the region above $\sim400$ GeV. With the additional
quark-gluon channel in the real radiation it is not surprising that the kinematics of the process changes compared to the 
leading order behavior, and this observable, being very inclusive in the final state probes the underlying kinematics of
the process.
One can expect also differences between signal and background in various angular distributions as in the case of the signal
the $b$-jets and the photons stem from the decay of a Spin-0 particle, whereas for the background processes the angular
correlations are different. Fig.~\ref{fig:dr} shows the $R$-separation between the two leading jets and the two photons 
respectively. Both distributions exhibit large corrections for small values for the separation with a minimum around $\pi$.
A similar behavior is also found for the azimuthal angle which is shown in Fig.~\ref{fig:dphi} for 
the leading jets (l.h.s) and the two photons (r.h.s). Also there one finds for both the jets and the photons the largest contributions
for small angles followed by a constant decrease. For $\Delta \phi \approx \pi$ the NLO result agrees with the LO result within
the theoretical uncertainty.

\subsection{Massive $b$-quarks}
\begin{figure}[t!]
  \centering
  \includegraphics[width=0.49\textwidth]{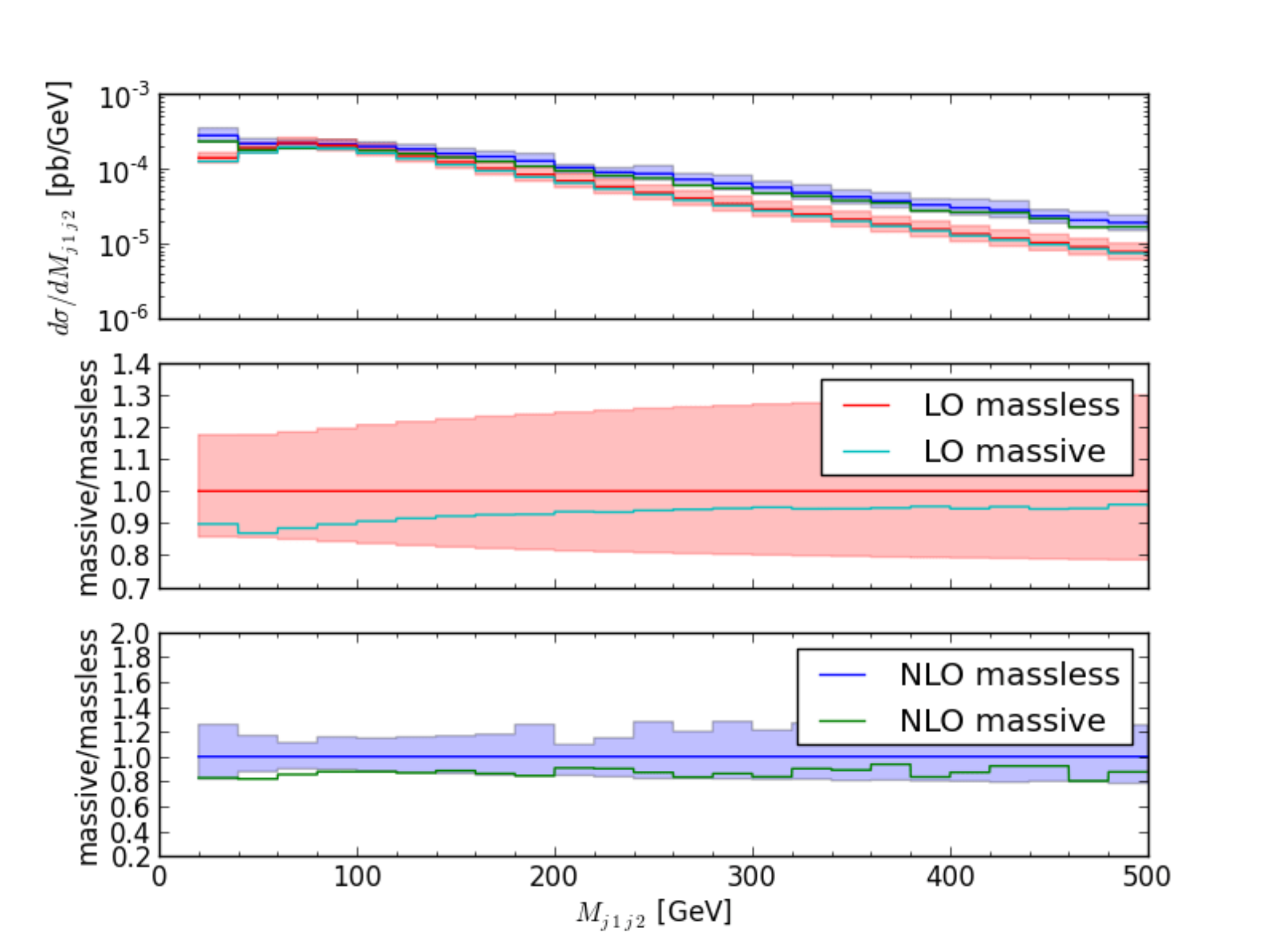}
  \hfill
  \includegraphics[width=0.49\textwidth]{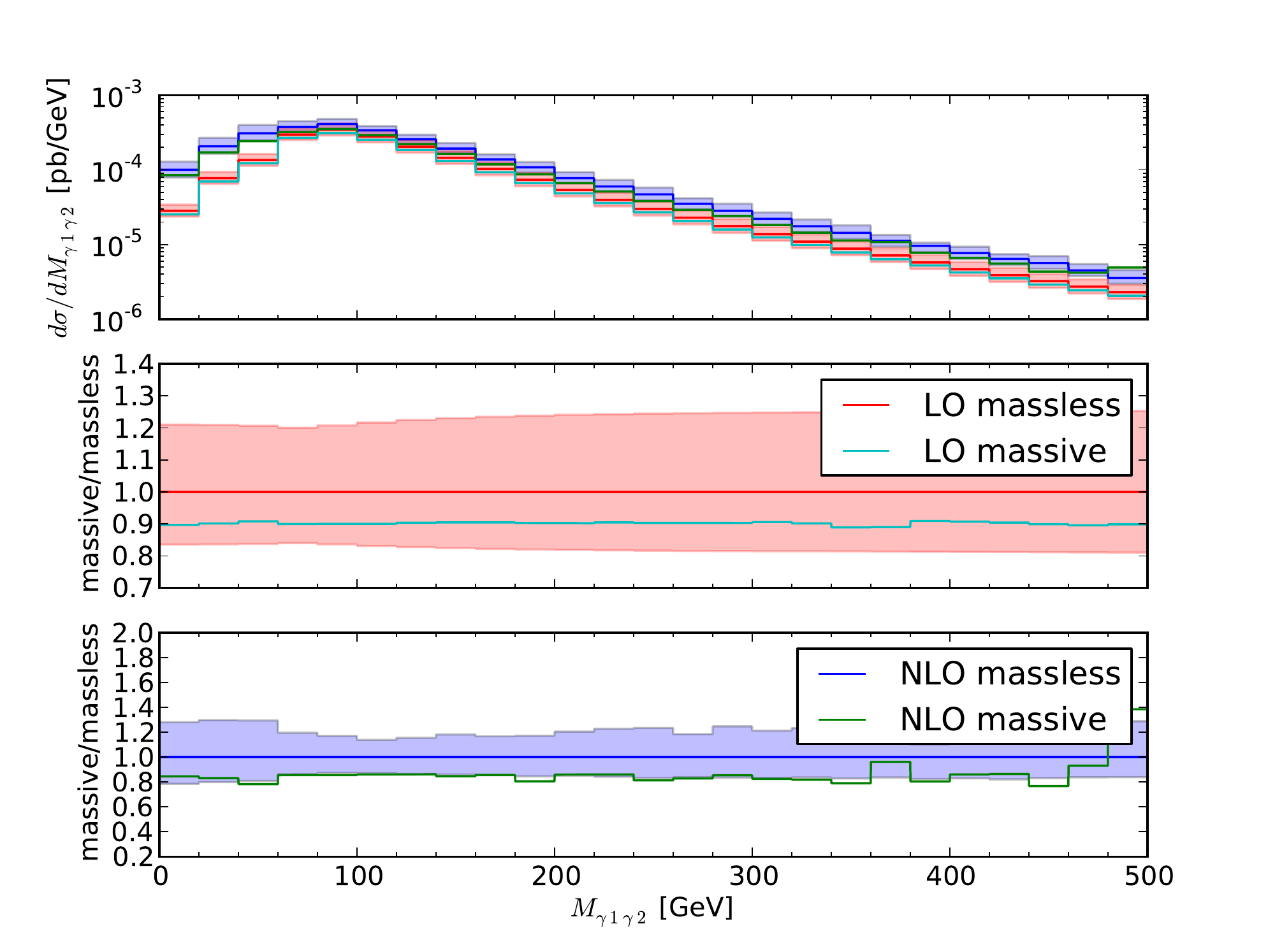}\\
  \includegraphics[width=0.49\textwidth]{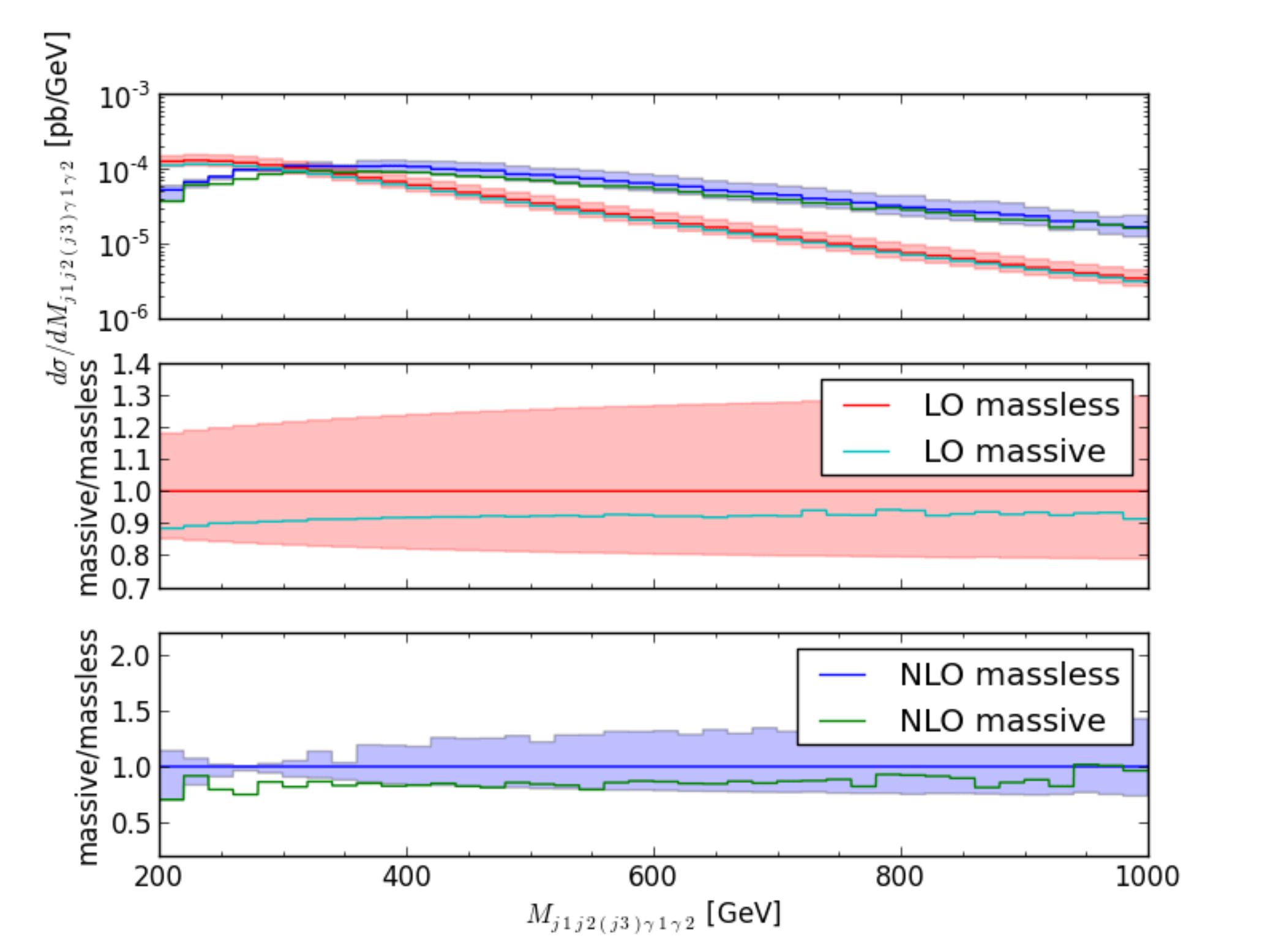}
  \caption{\label{fig:m_invar_massive} Invariant mass distribution for the two leading jets and the two photons (upper row)
  and for the total invariant mass (lower row) for massive
  $b$-quarks.}
\end{figure}

\begin{figure}[t!]
  \centering
  \includegraphics[width=0.49\textwidth]{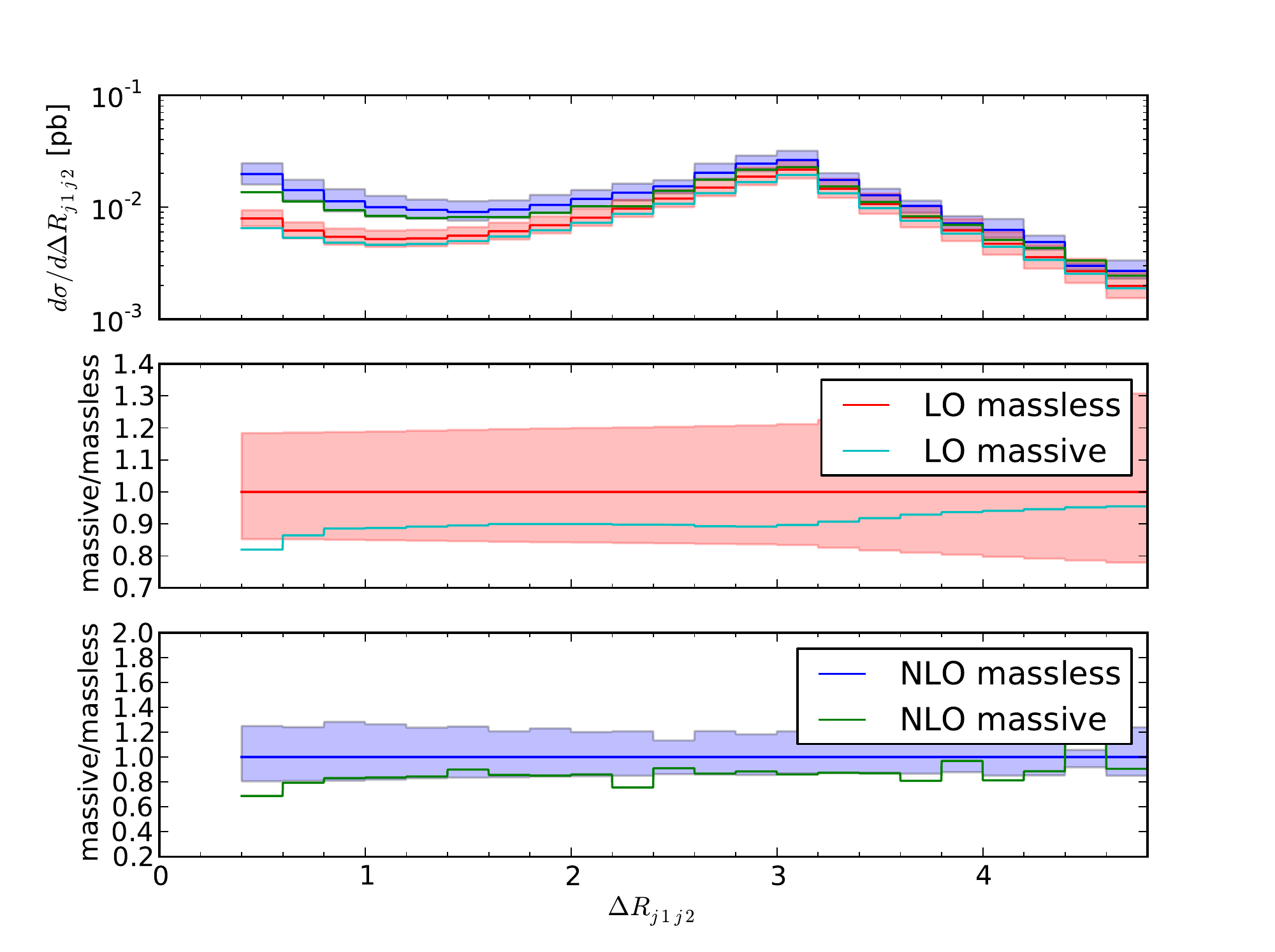}
  \hfill
  \includegraphics[width=0.49\textwidth]{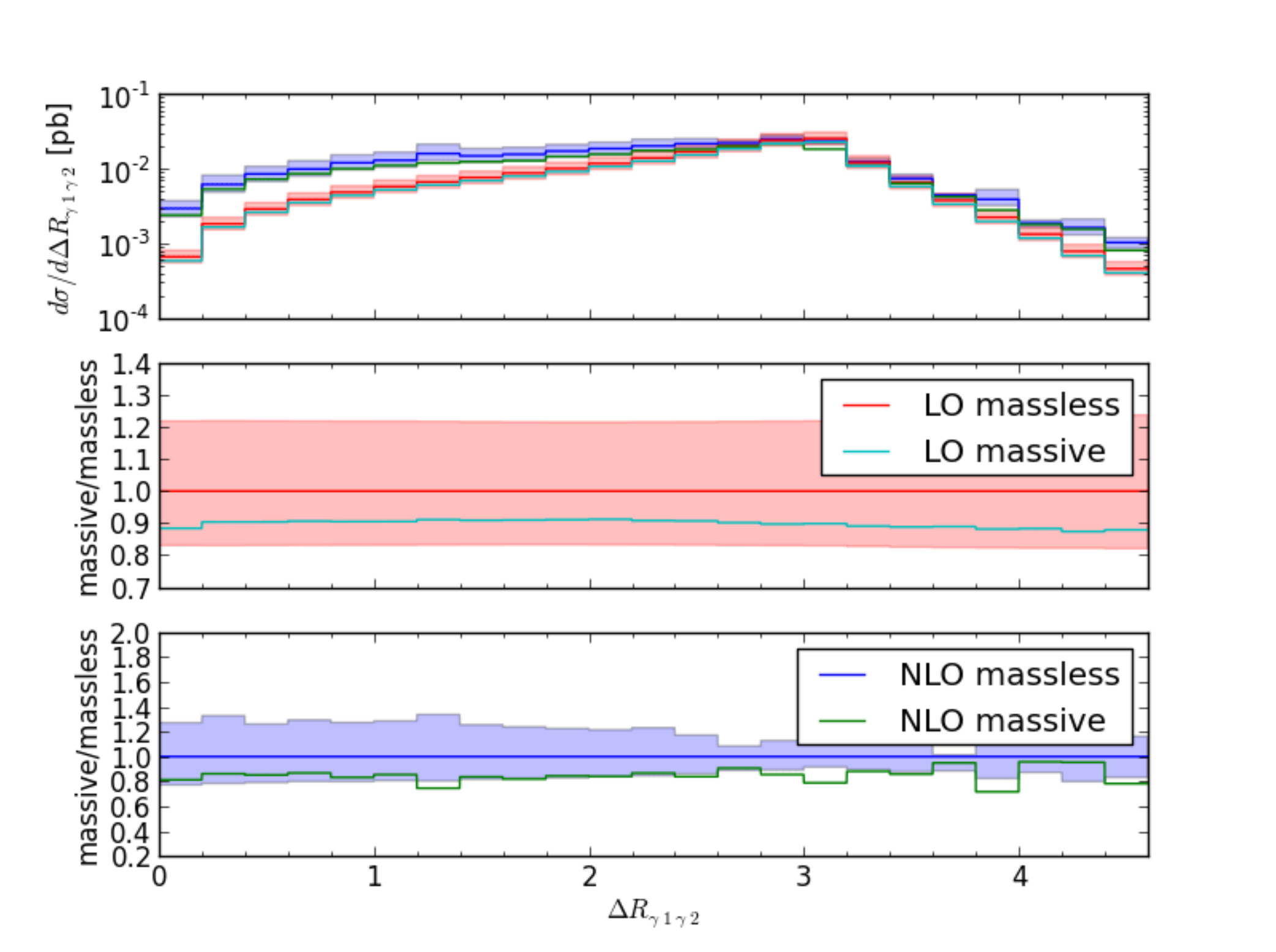}
  \caption{\label{fig:dr_massive} Angular separation between the two leading jets and the two photons for massive
  $b$-quarks.}
\end{figure}

In this section we will scrutinize the validity of treating the $b$-quark as a massless particle. For this purpose we set the
mass of the $b$-quark to its pole mass of $4.7$ GeV. For a consistent treatment we employ the four flavor scheme and use 
the CT10nlo\_nf4 pdf set. Table \ref{tab:xsec} shows the total cross sections for the central scale  at LO and NLO for massive 
quarks in direct comparison to the massless results. The massive LO order result is reduced by $\sim10\%$, at NLO 
the massive result is $\sim16\%$ smaller than the massless result. This reduces the $k$-factor by $\sim7 \%$. \\
At first it may seem unreasonable that the introduction of the $b$-mass does have such an influence on the result given that the 
mass is relatively small compared to all other scales in this process. However one should keep in mind that there are several 
effects that need to be taken into account. The biggest effect certainly comes from the change of the pdf set that comes along with a lower value of $\alpha_s$. $\alpha_s(M_Z)$ is $\sim 4.5 \%$ smaller in the massive case. This effect is the driving force
in the reduction of the cross section. In addition,  for this process the subprocesses with initial state $b$-quarks are enhanced due to t-channel like diagrams with the $b$-quark line going from initial to final state. These type of diagrams yield a large
contribution that enhances the importance of initial state $b$-quarks compared to the other sea-quark contributions.
This effect has also been observed in the context of multiple $b$-quark production \cite{Binoth:2009rv,Greiner:2011mp}
and also there the overall effect has been found to be large \cite{Bevilacqua:2013taa}. 
From comparing LO order results within the 4 flavor scheme for the massless and the massive case we estimate the pure
mass effect to contribute to $\sim 40 \%$ to the reduction of the cross section.\\
\begin{table}[h!]
\centering
\begin{tabular}{l|c|c}
$\mu=\mu_0$  & $m_b= 0$ GeV & $m_b= 4.7$ GeV \\ 
\toprule
$\sigma_{LO}$ [fb] & $38.62(2)$ & $34.83(1)$ \\
\midrule
$\sigma_{NLO}$ [fb] & $56.2(4)$ & $47.4(4)$ \\
\midrule
$K=\frac{\sigma_{NLO}}{\sigma_{LO}}$ & 1.46 & 1.36\\ 
\end{tabular}
\caption{\label{tab:xsec}Total cross sections at LO and NLO for the central scale and for massless and massive $b$-quarks.}
\end{table} 
It is now also important to investigate if and how big the massive $b$-quark will affect differential distributions. For simplicity
we present the massive results only for the central scale. The focus here is on the change of the shape caused by the mass 
effects and we assume that the theoretical uncertainty will be of a very similar size as for the massless case.  
\begin{figure}[t!]
  \centering
  \includegraphics[width=0.49\textwidth]{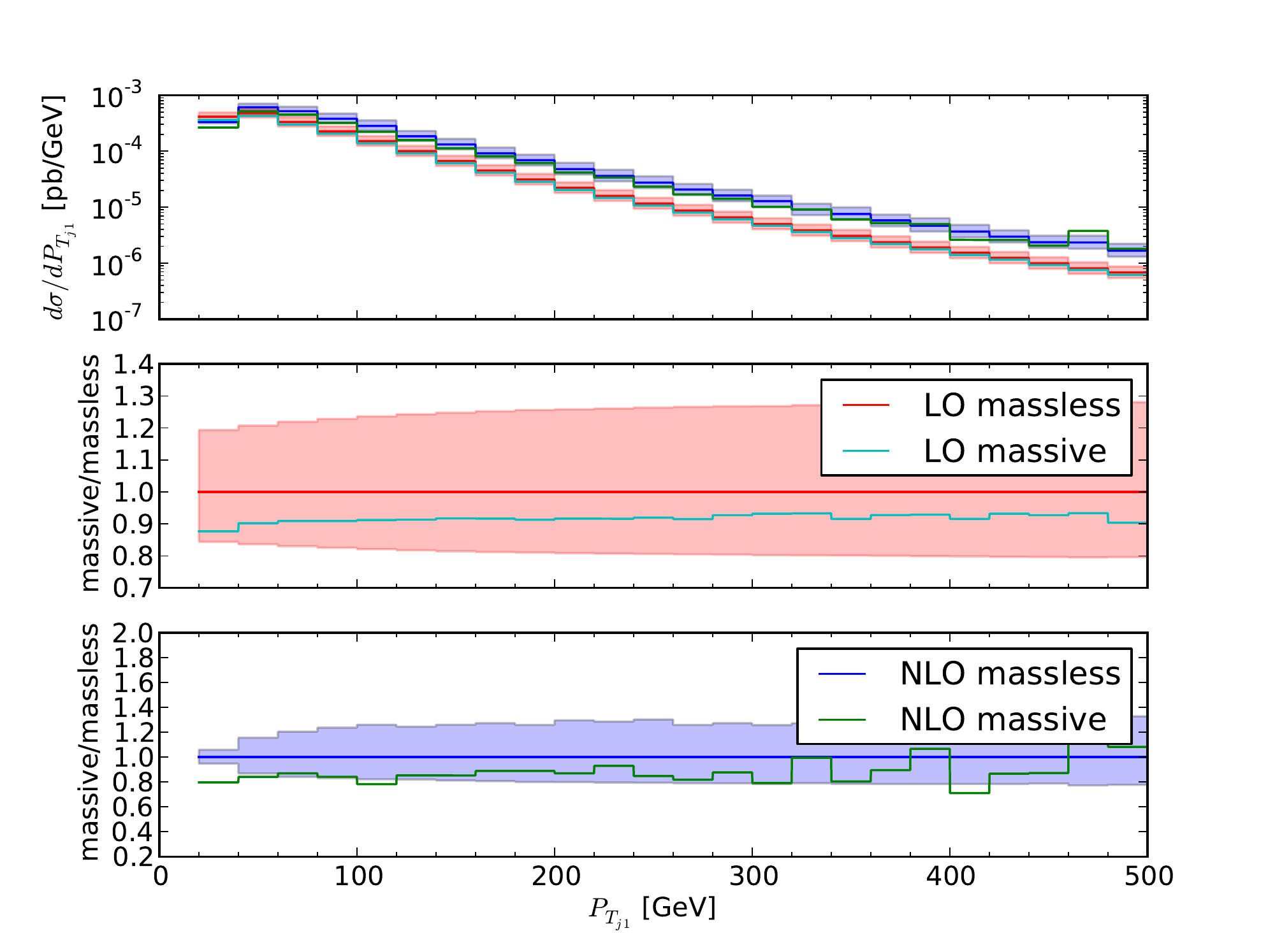}
  \hfill
  \includegraphics[width=0.49\textwidth]{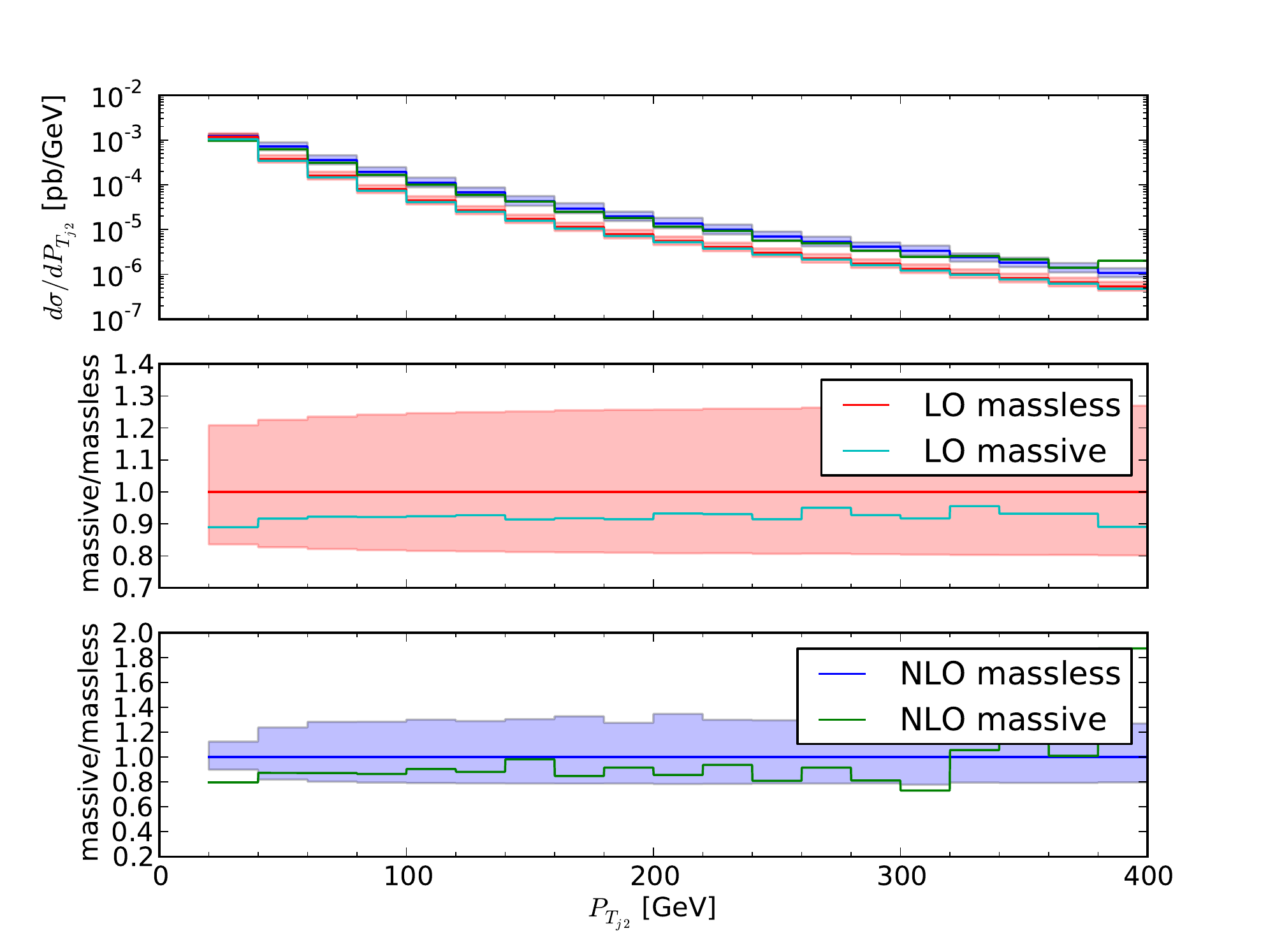}
  \caption{\label{fig:jet_pt_massive} Transverse momentum distribution of the two leading jets for massive $b$-quarks.}
\end{figure}
In Fig.~\ref{fig:jet_pt_massive} we show the transverse momentum distribution for the two leading jets. The upper
ratio plot shows the ratio of the massive LO contribution over the massless result, the lower ratio plot shows the
same for the NLO result. For comparison we also show the scale uncertainty for the massless case. 
The mass effects are dominated by the general decrease of the cross section in the massive case, but the differential
$k$-factor is flat to a quite good approximation and the central scale of the massive result is still within the uncertainty
band of the massless result except for the first bin where the uncertainty band becomes smaller. 
One can therefore conclude that the uncertainty from setting the mass to a non-zero value
is contained within the systematic uncertainty from scale variation.
\begin{figure}[t!]
  \centering
  \includegraphics[width=0.49\textwidth]{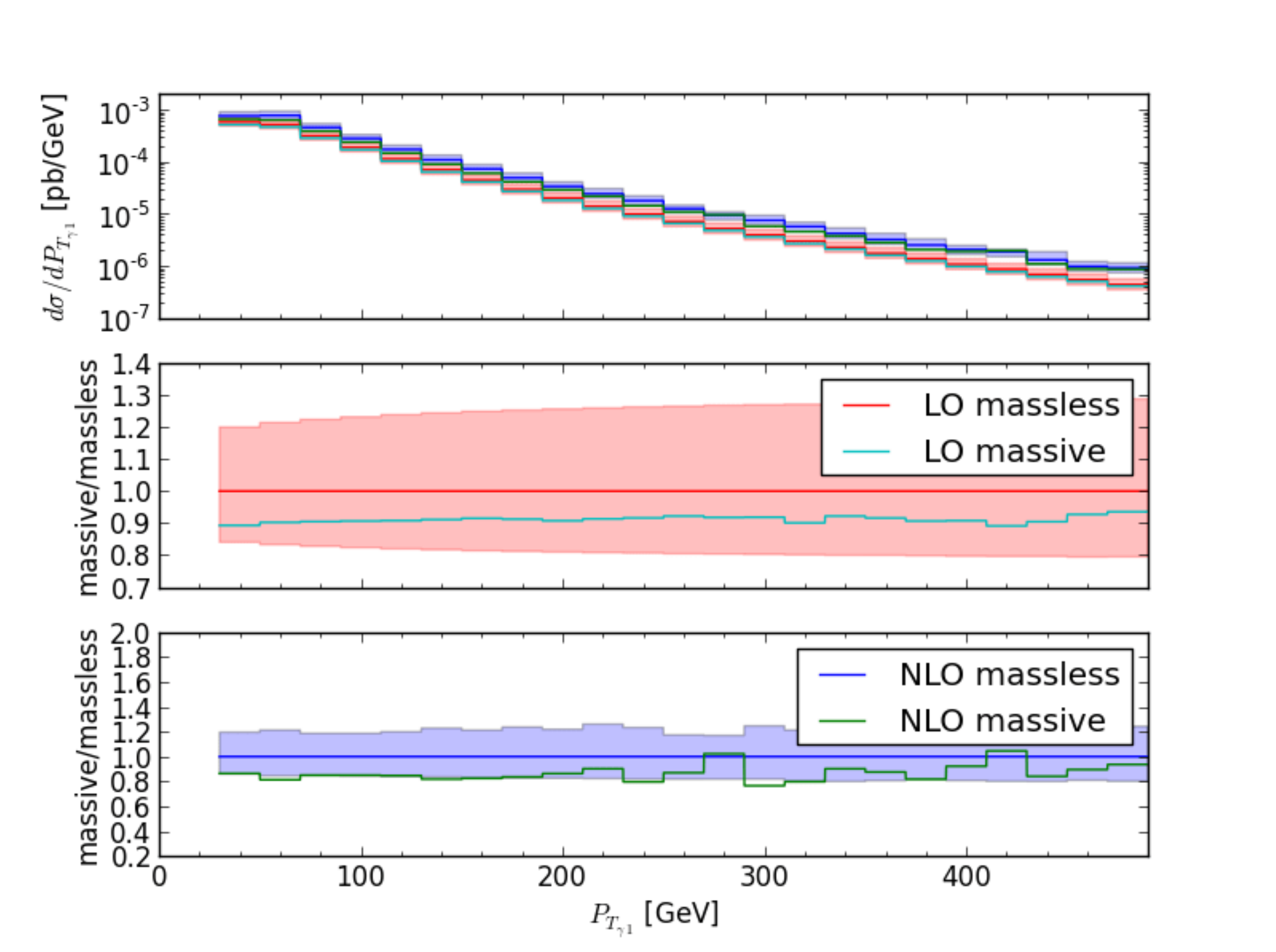}
  \hfill
  \includegraphics[width=0.49\textwidth]{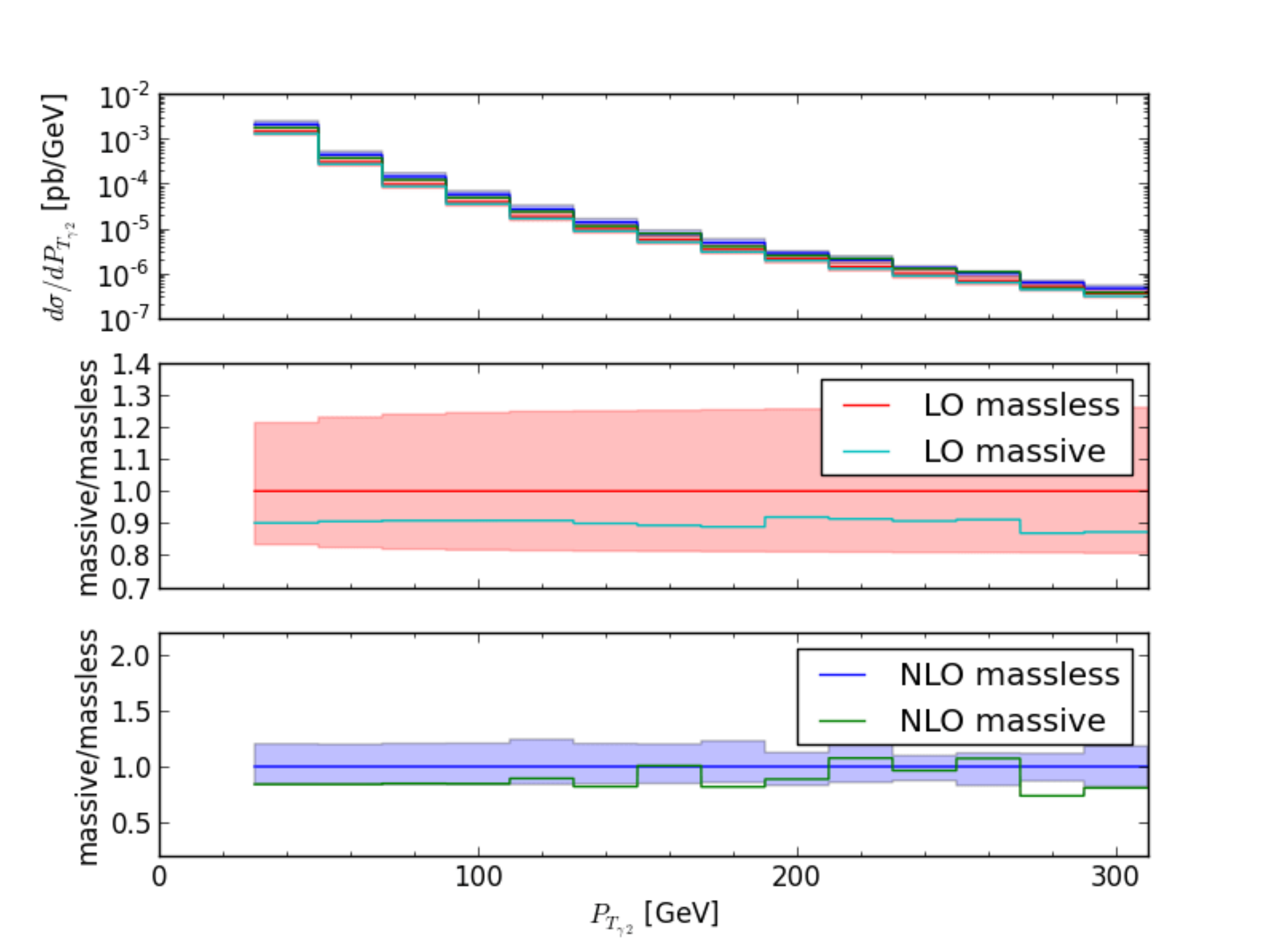}
  \caption{\label{fig:photon_pt_massive} Transverse momentum distribution of the two photons for massive $b$-quarks.}
\end{figure}
The transverse momentum distribution of the two photons shows exactly the same behavior as can be seen in
Fig.~\ref{fig:photon_pt_massive}. Also here the differential $k$-factor is flat to a good approximation and the massive
result is still in agreement within the uncertainty of the massless result. The same is also true for the invariant masses
of the two leading jets and the two photons as can be seen in Fig.~\ref{fig:m_invar_massive}. Also here the massive
result can be incorporated in the systematic uncertainty of the massless calculation. For the total invariant mass shown
in the lower row of Fig.~\ref{fig:m_invar_massive} the situation is a bit more special. Also here the mass leads to a flat
shift downwards, but the ratio plots shows that for the NLO result the error band becomes very small in the region between
$250-300$ GeV. The reason for this behavior is that the upper and the lower scale cross the central scale in that region
which makes the scale uncertainty vanish and leaving the massive result outside the estimated uncertainty.
This might also be interpreted such that for this observable our scale choice is not suitable to describe this 
particular observable and give a reliable estimation of the underlying uncertainties. 
In general on would expect that if the introduction of a massive $b$-quark yields to a shape distortion compared to the
massless case, then this should preferably show up in distributions that separate regions of low and high energy / transverse
momentum, such that there are regions where the $b$-mass becomes large compared to the other scales in the process. 
Distributions like the transverse momenta of the $b$-jets or the invariant mass of the dijet system seem to be the ideal
candidates. However as we have seen above, even in these distributions we do not observe a significant shape distortion
and the effects of the $b$-mass are essentially reduced to a global shift induced by the different value of $\alpha_s$.
It is therefore not surprising that also in angular distributions we do not observe a different pattern. We exemplify
this by showing the angular separation between the two leading jets and the two photons in Fig.~\ref{fig:dr_massive}.
As for the transverse momentum distribution of the jets one observes a small effect in the first bin where the massive result
is slightly below the uncertainty band of the massless result but also here the differential $k$-factor is flat over the whole range.
For the separation between the photons the situation is very similar with the massive result being at the lower end of the 
uncertainty band with an otherwise flat $k$-factor.\\
In summary, the inclusion of the $b$-mass has a substantial effect on the total cross section and on differential distributions.
However it leads just to a global shift towards smaller values largely caused by the 4 flavor pdf set and the smaller value
for $\alpha_s$.  But it does not lead to significant distortions of shapes of the differential distributions. A shift via a global 
$k$-factor would therefore be able to accurately describe the mass effect.

\section{Conclusions}
\label{sec:conclusions}
The measurement of the triple Higgs coupling is an essential ingredient to completely determine the structure of the 
Higgs potential and to answer the question whether the Higgs boson is in agreement with the prediction from the Standard
Model. The production of two Higgs bosons via gluon fusion yields the biggest contribution that includes the triple Higgs
vertex.\\
In this paper we investigated the background of one of the most import decay channels, where one Higgs would decay into
a $b\bar{b}$ pair and the other Higgs would decay into a pair of photons. We calculated 
the ${\cal O}(\alpha_s^2 \alpha^2)$ contribution at next-to-leading order
in QCD in the fully automated Sherpa + GoSam setup. We found large corrections due to new partonic channels opening up for the real emission contribution leading
to a tree-level like behavior of the cross section under variation of renormalization- and factorization scale. The inclusion
of NLO effects is therefore viable for a reliable theoretical prediction.\\
We also assessed the impact of a massive bottom quark. In a consistent treatment the inclusion of the mass comes along
with a 4 flavor scheme pdf set and therefore also the removal of subprocesses with initial state $b$-quarks. Altogether we
found a significant reduction of the cross section which however is largely caused by the pdf set and the smaller value of
$\alpha_s$. The actual mass only plays a minor role.  The massive result is still contained within the systematic
uncertainty of the massless one and the shapes of the differential distributions are unchanged to a good
approximation. This means that the mass effects can effectively be described by applying a global $k$-factor to the
massless results.

\section*{Acknowledgements}
We would like to thank Marek Schoenherr for his help with Sherpa and 
Thomas Gehrmann and Gudrun Heinrich for various useful discussions.
NG was supported by the Swiss National Science Foundation under contract
PZ00P2\_154829.

\providecommand{\href}[2]{#2}\begingroup\raggedright\endgroup

\end{document}